\documentclass{aa}
\usepackage[dvips]{graphics}
\usepackage{latexsym}
\def\kms{\,km\,s$^{-1}$}

\def\micron{\,$\mu$m}

\begin{document}
\thesaurus{20(11.06.2; 11.19.2; 13.09.1)}
\title{
Near-infrared observations of galaxies in Pisces-Perseus:
I. {\sl H}-band surface photometry of 174 spirals
\thanks{Based on observations at the TIRGO, NOT, and VATT telescopes.
TIRGO (Gornergrat, CH) is operated by CAISMI-CNR, Arcetri, Firenze. 
NOT (La Palma, Canary Islands) is operated by NOTSA, the Nordic Observatory
Scientific Association. VATT (Mt. Graham, Az) is operated by VORG, the
Vatican Observatory Research Group.}
}
\author{ G. Moriondo \inst{1} 
  \and C. Baffa \inst{2} 
  \and S. Casertano \inst{3}
  \and G. Chincarini \inst{4} 
  \and G. Gavazzi \inst{5} 
  \and C. Giovanardi \inst{2}
  \and L.K. Hunt \inst{6} 
  \and D. Pierini \inst{7} 
  \and M. Sperandio \inst{4} 
  \and G. Trinchieri \inst{8}
}
\institute{ Dipartimento di Astronomia e Scienza dello Spazio, Universit\`a di Firenze, Largo E. Fermi 5, I--50125 Firenze, Italy
\and
  Osservatorio Astrofisico di Arcetri, Largo E. Fermi 5, I--50125 Firenze,
  Italy
\and
  STECF Space Telescope Science Inst.,
  3700 San Martin Drive, Baltimore, MD 21218, USA
\and
  Osservatorio Astronomico di Brera, Via E. Bianchi 46, I--22055 Merate (CO), Italy
\and
  Dipartimento di Fisica, Sez. Astrofisica, Universit\`a di Milano,
  Via Celoria 16, I--20133 Milano, Italy
\and
  C.A.I.S.M.I C.N.R., Largo E. Fermi 5, I--50125 Firenze, Italy
\and
  Max-Planck-Inst. f\"ur Kernphysik, Bereich Astrophysik, Saupfercheckweg 1,
  Postfach 10 39 80, D--69029 Heidelberg, Germany 
\and
  Osservatorio Astronomico di Brera, Via Brera 28, I--20121 Milano, Italy
}
\offprints{gmorio\@arcetri.astro.it}
\mail{gmorio@arcetri.astro.it}
\date{Received ; accepted }
\titlerunning{NIR observations of spirals in PP}
\authorrunning{G. Moriondo et al.}
\maketitle
\begin{abstract}
We present near--infrared, $H$--band (1.65\micron), surface photometry of 174
spiral galaxies in the area of the Pisces--Perseus supercluster. The images, 
acquired with the ARNICA camera mounted on various telescopes,
are used to derive radial profiles of surface brightness, ellipticities,
and position angles, together with global parameters such as 
$H$-band magnitudes and diameters
\footnote{
Table 3 and Fig. 4 are only available in electronic
form at the CDS via anonymous ftp to cdsarc.u-strasbg.fr (130.79.128.5) or
via http://cdsweb.u-strasbg.fr/Abstract.html.
Radial profiles in tabular form and
images FITS files are also available upon request from gmorio@arcetri.astro.it.}. 
The mean relation between $H$-band isophotal diameter $D_{21.5}$ and the
$B$-band $D_{25}$ implies a $B-H$ color of the outer disk bluer than 3.5 ;
moreover, $D_{21.5}/D_{25}$ depends on (global) color and absolute 
luminosity. The correlations among the various
photometric parameters
suggest a ratio between isophotal radius
$D_{21.5}/2$ and disk scale length of $\sim 3.5$ and a mean disk central
brightness $\simeq 17.5 \,H$-mag arcsec$^{-2}$.
We confirm the trend of the concentration index $C_{31}$ with 
absolute luminosity
and, to a lesser degree, with morphological type.
We also assess 
the influence of non-\-axi\-sym\-me\-tric struc\-tures on the radial profiles
and on the derived parameters. 
\end{abstract}
\section{\protect\label{itrod}Introduction}
In this paper we present near--infrared (NIR) $H$-band images 
of a sample of 174 spiral galaxies located 
in the region of the Pisces-Perseus (PP) supercluster, the largest
low-redshift structure in the southern Galactic hemisphere.
Although a number of measured infrared magnitudes 
exist in the literature for some clusters
in this area (Aaronson et al. 1986),
this is the first large NIR imaging survey of this structure and,
together with the published surveys in the Coma supercluster
(Gavazzi et al. 1996a, c) and in Virgo (Boselli et al. 1997),
it makes up an extended database to investigate the properties of 
these most outstanding low-redshift superclusters and of their member
galaxies.
The data were collected to pursue several distinct lines of research:
to probe the internal extinction in spiral disks (Moriondo et al. 1998c);
to map peculiar motions in the supercluster (Baffa et al. 1993);
to study stellar populations and dark matter content (Moriondo et al. 
1998a,b). 

As detailed in the following section, 
the galaxies observed pertain to two distinct sets: the first one
was observed in the $H$-band only; the second one in three different 
passbands: $J, H,$ and $K$, when possible. 
We defer the publication and analysis of the colour images to a forthcoming
paper, and report here the $H$-band observations, that is the
passband with the largest wealth of  data, both in the literature and in
our present data set.
Although the resulting sample lacks completeness, 
the data are
remarkably homogeneous both because of the type of objects included and for the
uniformity in the observational and image reduction techniques. 
As it will be shown,
the sample 
comprises a fair representation of the bright disk galaxies
hosted in the entire supercluster. 
\section{Sample}
\subsection{\protect\label{select}Sample selection}
\setcounter{figure}{0}
\begin{figure}
\vspace{1.5cm}
\caption{Position of the sample objects in equatorial coordinates. 
In this equal-\-area projection, 
the clustering at $1^h 20^m$ and $33^\circ$ coincides with the
Pisces cluster (sample B).}
\label{sky-distr}
\end{figure}
\begin{figure}
\vspace{1.5cm}
\caption{Number of galaxies with apparent magnitude brighter than $m_B$
vs. $m_B$ ($B_T$ system). Error bars are $1\sigma$ uncertainties due to counting
statistics. The dotted
line, shown for comparison, has the slope 
0.6 expected for a uniform distribution in euclidean space; the position of the
line is arbitrary.}
\label{B-distr}
\end{figure}
The sample objects were selected from the catalogued
disk galaxies in the Pisces-Perseus supercluster area,
$22^h<R.A.<4^h$ and $0^\circ<\delta<45^\circ$, 
excluding obvious foreground members with $V_H<3000\,$\kms$\,$
and ellipticals.
\begin{figure}
\vspace{1.5cm}
\caption{Morphological distribution of the sample galaxies.
N is the number of objects per RH classification bin, T is the index of stage
along the Hubble sequence
from the RC3. Spiral types are also reported on top of each bin. The rightmost
bin, coded S, refers to objects classified S or SB without further specification.}
\label{T-distr}
\end{figure}
\addtocounter{figure}{+1}
To best fit the camera fields of view to be employed,
the sample was restricted to galaxies with optical major axis
$0\farcm 5<a<4\farcm 0$, and, among these, we further selected 
those with available good-quality 21-cm spectra (Giovanelli \& Haynes 1989 and references therein). These selection
criteria result into about 950 galaxies, mainly late spirals. 
From this set two distinct subsamples were extracted.
Sample A comprises about 150 galaxies, randomly selected to
cover the entire area
and inclination range, and with types Sb or later; for these
objects only $H$-band data were acquired. Sample B includes the 
50 cluster galaxies with distance $d_c < 5 \fdg 5$ from the
Pisces cluster center ($1^h 20^m, 33^\circ$),
with optical size $D_{25} \geq 0\farcm 8$,
and inclination $30^\circ < i < 75^\circ$;
for these objects, 
ranging in morphology from S0 to Sd, 
$J, H,$ and $K$-band images were acquired.

Due to malfunctions and bad weather, not all the sample objects
could be observed.
As a whole 177 objects were observed at least in one band (usually $H$),
147 of which belong to the supercluster sample (sample A)
and 35 to the cluster one (sample B);
five galaxies are common to both samples.
This paper presents data for the 174 galaxies imaged in the $H$ bandpass.
\subsection{Sample statistics}
The spatial distribution of the sample objects is depicted,
in an equal-\-area projection, in Fig. \ref{sky-distr}.
The region is boun\-ded to the east by the Zone of Avoidance
due to extinction in the Galactic plane.
Despite its scarcity, our sample delineates 
the overall structure of the supercluster as portrayed
by much richer sets (e.g. Giovanelli et al. 1986). The main
ridge appears clearly, spanning the entire R.A. range between 30$^\circ$
and 40$^\circ$ declination, as well as the southern extensions
down to the A400 and Pegasus clusters. Despite the presence of
several rich clusters in the area, the only clear clustering is
noted at the position of the Pisces cluster (NGC 507: $1^h 20^m \, \, \, 33^\circ$);
obviously this is due to the galaxies of sample B.

In Fig. \ref{B-distr} we plot the
cumulative distribution in photographic magnitude ($B_T$ system). 
When compared to the expectation for
a uniformly distributed set, the sample starts deviating and losing
completeness at $\sim 14~B$-mag ; there is also a notable a lack of bright
galaxies ($m_B < 13.0$). Both limits are mainly induced by
the constraints on the apparent size and only to a minor degree by
the characteristics of the UGC (Nilson 1973) and CGCG (Zwicky et al. 1961-68)
catalogues, from which the sample was originally drawn; 
the scarcity of bright objects also follows the
exclusion of foreground objects. 

The morphological distribution
is illustrated in Fig. \ref{T-distr}. The types are drawn from the
RC3 catalogue (de Vaucouleurs et al. 1991) where 
most of our sample galaxies have a detailed 
Revised Hubble (RH)
classification; only 16 are classified S or SB without further specification,
and only two non-UGC objects are not classified. 
As expected, due to the selection based on 21-cm observations, the sample
mainly consists of late spiral types and peaks around $T=4$, that is Sbc.
\section{The data}
\subsection{\protect\label{obs}Observations}
The sample galaxies were observed in the course of several observing campaigns,
from December, 1992 until February, 1997. Most of the data were collected 
at the TIRGO telescope (Zermatt, CH) in 1992, 1993, 1994 and 1997; some 
galaxies were observed at the NOT telscope (La Palma, Canary Islands) 
in 1995 and 
some at the VATT telescope (Mt. Graham, Az) in 1996.
The instrument used at all telescopes was the Arcetri NIR camera 
ARNICA, equipped with a 256$\times$256 pixel NICMOS3 detector
(Lisi et al. 1993, 1996; Hunt et al. 1996). The plate
scale was 0.97, 0.55, and 0.45 arcsec pixel$^{-1}$ at the
TIRGO, NOT and VATT respectively. 

All the galaxies were observed by alternating frames on the source and on
adjacent empty sky positions, integrating $\sim 1^m$ in each 
position. 
Typical total integration times on source were $\sim 5^m$ in all 
bands, yielding ($1\sigma$) surface magnitude limits of 
20.8, 21.3, and 20.6 $H$-mag arcsec$^{-2}$ at the TIRGO, 
NOT and VATT respectively (Hunt \& Mannucci 1998).
With a typical sky brightness of 14 $H$-mag arcsec$^{-2}$,
these limits correspond to $0.1 \, - \, 0.2$ \% of the ``background''.
\subsection{Data reduction}
All image reduction was performed with IRAF and the STSDAS
packages\footnote{IRAF is the Image Analysis and Reduction Facility
made available to the astronomical community by the National Optical
Astronomy Observatories, which are operated by AURA, Inc., under
contract with the U.S. National Science Foundation.
STSDAS is distributed by the Space
Telescope Science Institute, which is operated by the Association of
Universities for Research in Astronomy (AURA), Inc., under NASA contract
NAS 5--26555.}.
 
The flat fields for each source exposure were 
obtained by averaging the sky frames acquired immediately before and after it, 
upon removal from each sky frame of eventual stars in the field.
The use of medians of larger numbers of sky frames as flat fields (``superflats''),
despite their lower noise,
turns out to be effective only in case of exceptional stability of the
atmospheric emission. Due to the high sky brightness, the requirements 
on flatness for
mapping the outer regions of galaxies are quite strict. On average the
resulting flatness (low-spatial-frequency noise only) 
of our final images is about 0.01 \%
($1\sigma$), negligible relative to the high-spatial-frequency noise
which amounts to  $0.1 \, - \, 0.2$ \% as noted in Sect. \ref{obs}.
Such image quality allows a substantial improvement in signal-to-noise
of the elliptically-averaged profiles (see Sect. \ref{prof}).

The -- typically four -- flat-fielded source frames were subsequently 
cleaned for bad pixels, 
registered, and re\-scaled by an additive term to a common median level. 
They were then combined to obtain the median frame, after clipping 
deviant values in each position on the detector.
The final step was the subtraction of the background, 
which was estimated on the source frame itself upon
automatic exclusion of star and galaxy pixels.
\begin{table}
\caption{Extinction coefficients}
\begin{tabular}{lrrr}
\noalign{\smallskip}
\hline
\noalign{\smallskip}
  Telescope  & $J$      & $H$      & $K$     \\ 
\noalign{\smallskip}
\hline
\noalign{\smallskip}
         TIRGO &  0.083  &  0.029  & 0.063  \\
         NOT   &  0.057 &  0.000   & 0.042 \\
         VATT  &  0.067 &  0.007 & 0.040 \\
\noalign{\smallskip}
\hline
      &         &   &   \\
\end{tabular}
\protect\label{ext_coeff}
\end{table}
\subsection{\protect\label{calib}Photometric calibration}
A star from the list of northern standards by Hunt et al. (1998) was observed 
once about every hour in the three bands. In each band five frames were 
obtained with the star in different positions on the detector.
A flat field for each stan\-dard star position was obtained using the ``clip\-ped''
me\-dian of the remaining four positions. Stan\-dard-\-star frames were also 
cleaned for bad pixels, but not averaged together.
For every standard star observation, 
aperture photometry was performed in each position 
within a circle centered on the star, after subtracting the sky background. 
The radius of the aperture was typically 4 times the seeing FWHM; the background value 
was evaluated as the median in a circular annulus between radii 
5 and 8 FWHM.
The instrumental magnitudes were (to first order) 
corrected for atmospheric extinction 
using the coefficients reported in Table \ref{ext_coeff}
(from Hunt \& Mannucci 1998); units are mag airmass$^{-1}$. 
An average zero point was then computed for each night of observations
in every band and used to calibrate the relative galaxies.
\begin{table*}
\begin{flushleft}
\setcounter{table}{1}
\caption{Summary of observations}
\begin{tabular}{cccccl}
\noalign{\smallskip}              
\hline                           
\noalign{\smallskip}            
Telescope & Date & Zero Point &                               
$\sigma_{ZP}$ & FWHM & \multicolumn{1}{c}{Objects observed} \\           
 & [yy/mm/dd]  & [H mag] &        
[H mag] & [arcsec] & \\                                               
(1) & (2) & (3) &         
(4) & (5) & \multicolumn {1}{c}{(6)} \\        
\noalign{\smallskip}                                                             
\hline                                                                          
\noalign{\smallskip}                                                           
\vspace{+0.1cm}
T & 92/12/27 & 19.47 & 0.08 &  2.3 &  UGC: 1205, 1633 \\                     
\vspace{+0.1cm}
T & 93/10/16 & 19.51 & 0.04 &  2.1 &  UGC: 732, 835, 975 \\                     
\vspace{+0.1cm}
T & 93/10/18 & 19.26 & 0.20 &  2.6 &  UGC: 646, 820 \\                     
\vspace{+0.1cm}
T & 94/09/25 & 19.74 & 0.05 &  3.5 &  UGC: 673, 919 \\                     
\vspace{+0.1cm}
T & 94/09/26 & 19.76 & 0.08 &  2.7 &  UGC: 690, 11931, 12527 \\ 
\vspace{+0.1cm}
T & 94/09/27 & 18.93 & 0.27 &  1.8 &  UGC: 12098, 12103, 12234, 12614 \\    
\vspace{+0.1cm}
T & 94/09/29 & 19.76 & 0.11 &  2.1 &  UGC: 98, 100, 438, 484, 725, 12821, 12915 \\
\vspace{+0.1cm}
T & 94/11/16 & 19.83 & 0.02 &  3.4 &  UGC: 1688, 1792 \\                     
T & 94/11/18 & 19.82 & 0.03 &  2.6 &  UGC: 14, 57, 214, 230, 420, 463, 493, 11973,\\
\vspace{+0.1cm}
  &          &       &      &      &  12100, 12137, 12173, 12586, 12618 \\
T & 94/11/21 & 19.84 & 0.06 &  1.7 &  UGC: 89, 365, 798, 886, 911, 937, 962, 1094,\\
  &          &       &      &      &  1259, 1497, 1629, 2094, 2241, 12039, 12108,\\
\vspace{+0.1cm}
  &          &       &      &      &  12230, 12378, 12486, 12539, 12610, 12780 \\
T & 94/11/24 & 19.81 & 0.05 &  2.2 &  UGC: 16, 60, 301, 940, 1238, 1250, 1292, 2042,\\
\vspace{+0.1cm}
  &          &       &      &      &  2109, 2178, 2200, 12808, 12834 \\         
T & 94/11/25 & 19.85 & 0.04 &  2.1 &  UGC: 728, 796, 857, 906, 909, 1013, 1302, 1376,\\
  &          &       &      &      &  1411, 1768, 2586, 2604, 2608, 2627, 2655, 2658,\\
\vspace{+0.1cm}
  &          &       &      &      &  2885, 2931; CGCG~418-002 \\ 
T & 94/11/26 & 19.82 & 0.03 &  2.4 &  UGC: 448, 833, 1005, 1033, 1034, 1100, 1111, 1291, \\
  &          &       &      &      &  1582, 1626, 1835, 2048, 2142, 2156, 2353, 2435,\\
  &          &       &      &      &  2548, 2810, 11897, 12153, 12181, 12199, 12250,\\ 
\vspace{+0.1cm}
  &          &       &      &      &  12286, 12379, 12598, 12666, 12667; CGCG~503-007\\
T & 94/11/27 & 19.00 & 0.21 &  2.0 &  UGC: 1210, 1349, 1695, 1935, 2134, 2204, 2223, \\
\vspace{+0.1cm}
  &          &       &      &      &  2303, 2368 \\                      
N & 95/09/04 & 21.12 & 0.03 &  1.1 &  UGC: 26, 74, 114, 311, 414, 462, 714, 927, 1089, \\
\vspace{+0.1cm}
  &          &       &      &      &  1350, 1355, 1395, 1451, 1579, 1805, 2258, 2264 \\
\vspace{+0.1cm}
N & 95/09/05 & 21.10 & 0.04 &  1.1 &  UGC: 697; CGCG: 501-091, 502-085, 502-097, 521-038 \\                    
\vspace{+0.1cm}
N & 95/09/08 & 21.11 & 0.05 &  1.4 &  UGC~745 \\                    
\vspace{+0.1cm}
N & 95/09/09 & 21.06 & 0.04 &  0.9 &  UGC~738, CGCG~502-021 \\ 
\vspace{+0.1cm}
V & 96/11/28 & 20.41 & 0.05 &  1.2 &  CGCG~502-032 \\                    
\vspace{+0.1cm}
V & 96/12/01 & 20.43 & 0.01 &  1.8 &  UGC: 77, 776, 12776 \\                    
\vspace{+0.1cm}
V & 96/12/02 & 20.45 & 0.02 &  1.4 &  UGC: 106, 380, 1556, 1564, 1577 \\
\vspace{+0.1cm}
V & 96/12/03 & 20.46 & 0.02 &  1.9 &  UGC: 800, 831 \\                
\vspace{+0.1cm}
V & 96/12/04 & 20.35 & 0.04 &  0.9 &  UGC~710 \\                   
\vspace{+0.1cm}
T & 97/02/03 & 19.78 & 0.02 &  2.3 &  UGC: 1887, 1937, 2617; CGCG~521-068 \\
\vspace{+0.1cm}
T & 97/02/04 & 19.80 & 0.07 &  1.7 &  UGC: 1131, 2185 \\
\noalign{\smallskip}
\hline             
  &          &       &      &      &   \\  
\end{tabular}                                                                   
\end{flushleft}                                                                
\protect\label{logbk}
\end{table*}                                                                  
All the data presented here, magnitudes, surface magnitudes, and diameters,
take into account only correction for atmospheric extinction;
no correction has been applied for Galactic and internal extinction nor
for the effects of redshift (K correction, and angular size - $z$ relation).
A summary log of the obervations is presented in Table 2, where
for each observing night:
{\it Column 1} indicates the telescope used (N for NOT, T for TIRGO, V for VATT);
{\it Column 2} is the date of the following day;
{\it Column 3} is the nightly averaged zero point in the $H$ band;
{\it Column 4} is the $1\sigma$ uncertainty on the calibration as derived
by the set of standard-star observations for the night;
{\it Column 5} is the average FWHM of the seeing measured on the standard-star
images; 
{\it Column 6} is the list of sample galaxies observed that night.

$H$-band multiaperture photometry of 11 of the sample galaxies is 
available from Bothun et al. (1985), 
who used a single-element, InSb photometer; the galaxies are UGC 646, 673,
732, 820, 919, 927,
1013, 1033, 1094, 12486, 12359. 
Similar data are also available from Baffa et al. (1993) for UGC 57,
60, 98, 975, 1094, 1100, 1302, 1411, 2548, 12527, 12666.
In order to compare 
these with our results, we have performed multiaperture photometry on
our images using their aperture values, for a total
of 24 measurements to compare with Bothun et al. and 55 with
Baffa et al. . The mean difference between the results by Bothun et al. and ours is
$+0.04 \pm 0.08$ mag (standard deviation of the set, not of the mean),
a quite comfortable result; the comparison with Baffa et al. yields
a difference of $+0.07 \pm 0.14$ mag.
In neither case is a systematic difference 
clearly discernible\footnote{It may be noted that their
observations were performed with a chopping technique and on-line sky
subtraction. Due to the limited chopping throw, it is possible that 
external regions of the galaxy, as well as faint objects nearby,
are included in the sky measurement with a consequent
underestimate of the source flux within the aperture.
Because of this, and of the difficulty of centering faint objects within the
aperture diaphragms, single-photometer measurements are expected
to be slightly fainter than those acquired by panoramic detectors.}.
\subsection{\protect\label{prof}Brightness profiles}
For each of the final galaxy images the coordinates of the center
were determined by fitting a 
Gaussian to the center. Brightness profiles were then extracted
by fitting elliptical contours of increasing galactic radius, keeping the center
position fixed, with surface brightness $\mu_H$, ellipticity $\epsilon=1-b/a$, and position angle $PA$ as free 
parameters. The profiles were sampled for increasing values 
of the major semiaxis $a$, from 1 to 10 pixels in steps of 1 pixel,
and then outward with a $10\%$ geometric progression $a_{i+1}=1.1 \times a_i$.
Regions of the image containing close companions or foreground stars were edited out
from the fitted area. 

We also provide an estimate of the seeing FWHM for each final image
of a galaxy.
This was determined in 
most cases using stars in the field of
the galaxy image. When this was not possible, stars were selected in the 
images immediately before or after the one considered\footnote{Obviously  such
estimates, given in Table 3,
do not usually coincide with the nightly averaged FWHM from standard
stars listed in Table 2.}.
The resulting profiles for $\mu_H$, $\epsilon$, and $PA$ are reported
to the right of the corresponding image in Fig. 4.
\section{\protect\label{results}Results}
The global photometric parameters are summarized in Table 3, whose
entries are as follows.

\noindent
{\it Column 1}: UGC number or, if missing, CGCG
number;

\noindent
{\it Column 2}: Other common names;

\noindent
{\it Column 3 and 4}:  Equatorial J2000 coordinates in standard units 
({\sl hh mm ss.s, dd mm ss});

\noindent
{\it Column 5}: Revised Hubble morphological type from the RC3;

\noindent
{\it Column 6}: Total $B$ magnitude or, if missing, 
photographic magnitude reduced to the $B_T$ system, from 
the RC3;

\noindent
{\it Column 7}: Isophotal optical size, major ($D_{25}$) and minor axes in
arcmin at 25 $B$-mag arcsec$^{-2}$, from the RC3;

\noindent
{\it Column 8}: Heliocentric systemic velocity in km s$^{-1}$ as
listed in NED\footnote{This research has made use of the 
NASA/IPAC Extragalactic Database
(NED) which is operated by the Jet Propulsion Laboratory, California
Institute of Technology, under contract with the National
Aeronautics and Space Administration}, 
or, if missing, from private archives;

\noindent
{\it Column 9}: ``Total'' $H$ magnitude ($H_T$) within a circle of aperture
$D_{25}$;

\noindent
{\it Column 10}: Isophotal $H$ magnitude ($H_{21.5}$) within the elliptical isophote at 21.5
$H$-mag arcsec$^{-2}$;

\noindent
{\it Column 11}: Major axis ($D_{21.5}$) in arcmin of the elliptical isophote at 21.5
$H$-mag arcsec$^{-2}$;

\noindent
{\it Column 12}: Effective diameter ($D_e$) in arcmin; this is the major axis
of the elliptical isophote containing half of the flux corresponding to 
$H_T$;

\noindent
{\it Column 13}: Concentration index $C_{31}$, defined as the ratio between the
major axes of the ellipses enclosing 75\% and 25\% of the flux corresponding
to $H_{21.5}$;

\noindent
{\it Column 14}: Ellipticity $\epsilon=1-b/a$ of the outer elliptical isophotes;

\noindent
{\it Column 15}: Position angle ($P.A.$) of the outer elliptical isophotes, computed
Eastward from North;

\noindent
{\it Column 16}: Telescope of observation: TIRGO (T), NOT (N),
and VATT (V);

\noindent
{\it Column 17}: Estimate (FWHM) of the seeing disk of observation in arcsec,
see Sect. \ref{prof}.

\noindent
{\it Column 18}: Notes from the catalogues.

\begin{figure*}
\vspace{1.5cm}
\caption{The difference between the 
isophotal magnitude $H_{21.5}$ and the total magnitude
$H_T$ vs. $H_{21.5}$ (left panel), and vs. 
the average surface magnitude $\langle\mu_{21.5}\rangle$ (right panel). 
In both panels open symbols are used
when the radial brightness profile extends to levels 
fainter than 21.5 (interpolation); solid symbols
are used otherwise (extrapolation). In the left panel the dotted
lines are the loci of constant $\rho$, the ratio 
of the isophotal radius $D_{21.5}/2$ to
the disk exponential folding length $r_d$.
In the right panel the dotted lines are loci of
constant $\mu(0)$, the face-on disk surface brightness whose
value in $H$-mag arcsec$^{-2}$ is reported on top of each line.}
\label{h_h}
\end{figure*}
\subsection{\protect\label{magnitudes}Magnitudes}
Two magnitudes are here reported for the imaged galaxies: a total value
$H_T$ and an isophotal one $H_{21.5}$.

The total magnitude $H_T$ measures the
flux contained within a circular aperture the size of the optical
diameter $D_{25}$.
For our images this is always an extrapolated value and is computed with
a procedure similar to the one outlined in Gavazzi \& Boselli (1996) 
and Gavazzi et al. (1996a), although the values here are not corrected for
extinction and redshift. 
We estimate the average $1\sigma$ accuracy of $H_T$ to be $\sim 0.15$
mag, half the error being contributed by noise and calibration and half
by uncertainty in the extrapolation.

\begin{figure*}
\vspace{1.5cm}
\caption{The NIR isophotal diameter $D_{21.5}$ vs. the optical
size $D_{25}$ from the RC3. In the left panel only 
$D_{21.5}$ values derived by interpolation are reported and the sample objects have been grouped into
three classes according to the (optical) axial ratio $b/a$, and plotted
with differerent symbols. In the right panel extrapolated and
interpolated $D_{21.5}$ values are shown with different symbols.
The dashed lines are 
for the case $D_{21.5}=D_{25}$ and the dotted ones for $D_{21.5} = 0.89 D_{25}$,
the average slope for the sample.}
\label{radii}
\end{figure*}
The isophotal magnitude $H_{21.5}$ is derived by integrating 
the surface brightness radial profile from the center out to the 
elliptical isophote at 21.5 $H$-mag arcsec$^{-2}$.
In some cases, due to insufficient
field of view, or to a particularly noisy background, or to strong asymmetries,
we were not able
to fit elliptical contours down to such brightness levels. In these cases we
provide an estimate of $H_{21.5}$ obtained by (exponential)
extrapolation of the outer profile;
extrapolated values are enclosed in parenthesis and constitute roughly
10\% of the total. 
In practice, the extrapolation was performed by fitting a weighted
linear regression to three outermost points of the surface-magnitude
radial profile. In the few cases where such regression was 
deemed not satisfactory, the procedure was repeated with the 
6 outermost points of the profile.
We estimate the average $1\sigma$ accuracy of $H_{21.5}$ to be $\sim 0.08$
mag in case of interpolation and twice as much for the extrapolated values.
Again, such values are only corrected for atmospheric extinction.

As for the relation between the two types of magnitudes, we find
$\langle H_{21.5} - H_T \rangle = 0.20$, exactly what was
found for the sample in Gavazzi et al. (1996a).
The relation between the two $H$ magntudes is illustrated in Fig. \ref{h_h},
where their difference is plotted vs. the magnitude itself and 
vs. the average $H$ surface magnitude. Given the already quoted
accuracies for the two magnitudes, the distribution appears to consist
of a normal range, for $H_{21.5}-H_T < 0.3$ mag,
and by a deviant tail for the higher values. Such large differences
are partly due to the inclusion of some faint spurious objects,
such as foreground stars superimposed on the outer disk.
In general, the points for the extrapolated values, that is
when also $H_{21.5}$ had to be estimated by extrapolation
of the brightness profile, are distributed similarly to the 
others, which implies that, as expected, most of the variance is
contributed by $H_T$. A significant correlation is evident
between $(H_{21.5}-H_T)$ and $H_{21.5}$ itself;
upon inclusion of all points, the slope
is $0.050\pm 0.014$. 
A similar and tighter trend is detected in the dependence on
$\langle\mu_{21.5}\rangle$,
the average surface magnitude within the isophotal elliptical
contour at 21.5 $H$-mag arcsec$^{-2}$ (right panel). 
Such correlations are likely due to the narrow range of 
apparent diameters of the sample, see Sect. \ref{select}.
This selection criterion causes the faintest galaxies to be
often those with fainter surface brightness (and lower inclination)
and, consequently, with smaller isophotal size and fainter 
isophotal magnitude.

We conclude that the
accuracy in estimating our $H$ magnitudes,
especially the total magnitudes $H_T$,  
degrades for the faintest galaxies
of the sample, and that an appreciable part
of the error is systematic in the sense of yielding
too bright $H_T$ values for fainter $H_{21.5}$ and/or
$\langle\mu_{21.5}\rangle$. 
Such a trend spells a word of caution for the use of
heavily extrapolated magnitudes in, say, distance
measurements such as the Tully-Fisher relation. 

Since the difference $H_{21.5}-H_T$ is ideally determined only
by the outer disk, we note that, for an exponential disk with folding
length $r_d$, such difference depends only on 
$\rho=(D_{21.5}/2)/r_d$, the ratio between the isophotal radius and the 
folding one :
$$H_{21.5}-H_T = 2.5 \, \log_{10}\left[\frac{1}{1-e^{-\rho}\,(1+\rho)}\right]\,\, ,$$
if $H_T$ is identified with the total, asymptotic
magnitude of the disk.
Also, if $\mu(0)$ is the disk central surface magnitude :
$$\left[\,\langle \mu_{21.5} \rangle - \mu(0)\,\right]-(H_{21.5}-H_T)=5\,\log_{10}\left(\frac{\rho}{\sqrt{2}}\right)$$
In the left panel of Fig. \ref{h_h}, the dashed lines 
are curves of constant $\rho$, while in the right panel they are
curves of constant $\mu(0)$. The average (outer) disk appears to have an
isophotal radius $D_{21.5}/2 \simeq 3.5\, r_d$, with a central brightness
$\mu(0) \simeq 17.5$ $H$-mag arcsec$^{-2}$.
\subsection{Diameters}
The $D_{21.5}$ diameter, reported in Col. 7 of Table 3, is the major axis
of the elliptical isophote at 21.5 $H$-mag arcsec$^{-2}$. 
This is provided, in arcmin, for 
all the objects observed in the $H$ band. 
As for the isophotal magnitudes, in some cases 
$D_{21.5}$ had to be measured by (exponential)
extrapolation of the outer profile;
extrapolated values are enclosed in parenthesis. 
The average $1\sigma$ accuracy of $D_{21.5}$ is usually about 5"
and three times worse in case of extrapolation.
 
Figure \ref{radii} illustrates the comparison between our $D_{21.5}$ and the $B$-band
$D_{25}$ from the RC3.
In the left panel only $D_{21.5}$ values obtained by interpolation are
reported and the data set has been divided into three bins of (optical) axial ratio $b/a$;
the bin boundaries are those which result in bins with equal number
of objects. Within the
uncertainties, the relation $D_{21.5}$ vs. $D_{25}$ does not deviate from 
linearity. As a whole we find $D_{21.5} = 0.89\, D_{25}$, 
which implies that the 21.5 $H$-mag~arcsec$^{-2}$
is not as deep as the standard 25.0 $B$-mag~arcsec$^{-2}$; in other words,
the $(B-H)$ colour of the outer galaxy regions is bluer, on average,
than 3.5 (see also de Jong 1996). 
As for the comparison of the different inclination bins,
we find no significant difference or trend: $D_{21.5} = 0.88\, D_{25}$ for 
$b/a < 0.6$, $ D_{21.5} = 0.85\, D_{25}$ for $0.6 < b/a < 0.8$, and
$D_{21.5} = 0.91\, D_{25}$ for $b/a > 0.8$.
If any, the effects of internal extinction are not noticeable on
this relation, which confirms the overall transparency of the outer
disk.
\begin{figure}
\vspace{1.5cm}
\caption{The ratio between the isophotal $H$-band diameter
$D_{21.5}$ and the $B$-band $D_{25}$ vs. the total $B-H$ colour $m_B-H_T$
(upper panel), and the colour-magnitude relation for the present sample (lower
panel).
Only the galaxies with $D_{21.5}$ measured by interpolation are reported.
The line shown for comparison illustrates the behaviour
of an exponential disk with
fixed central surface magnitudes (see text). 
}
\label{color_diam}
\end{figure}
A last comment regards the rather large scatter about the average regression
with $1\sigma \simeq 30\%$. 
This is true, and approximately constant, over the whole range
of apparent size and does not depend on particularly deviant cases;
as shown in the right panel of Fig. \ref{radii}, 
it actually remains the same upon exclusion of the extrapolated $D_{21.5}$ values. 
The scatter can be attributed to the different methods of measurement: 
an objective
estimate from the elliptically averaged profile in the case of $D_{21.5}$,
and inspection of the 2D image for $D_{25}$. Especially in the case of late
spirals, inspection of $B$-band plate material is strongly
affected
by spiral structure and by the presence of outer H{\small II} complexes.
Further discussion of this issue will be found in Sect. \ref{ell_sect}.

The ratio between the isophotal optical and NIR diameters is 
found to be a weak function of the galaxy colour. This is illustrated in 
the upper panel of Fig. \ref{color_diam} where the ratio is plotted 
against the total $B-H$ index, $m_B-H_T$. 
The solid curve represents an exponential disk with a central 
$B-H=3.5$ mag and different scale lengths in the two bandpasses; 
for this special value of the central colour ($3.5=25-21.5$) 
the ratio $D_{21.5}/D_{25}$
is equal to the scale lengths ratio, and the total colour only depends
on this ratio.
Due to
the well-known colour-magnitude relation (Tully et al. 1982),
the dependence on colour also implies a certain
dependence on the absolute luminosity; the colour-magnitude relation 
for our sample is shown in the lower panel of Fig. \ref{color_diam}.
The indicative absolute magnitudes are computed assuming
a redshift distance with $H_o=100$ km s$^{-1}$ Mpc$^{-1}$ and upon reduction
of the velocity to the Local Group centroid according to RC3 (no infall correction);
assuming a solar absolute magnitude of 3.39 $H$-mag, as in Gavazzi et al. (1996b),
the mean value of about -22.5 mag is equivalent to $2.3 \times 10^{10}$ L$_{\sun}$.
No clear correlation was instead found between $D_{21.5}/D_{25}$ and the
apparent parameters (magnitude, size, inclination) so that our $D_{21.5}$
estimates appears to be generally free of measurement biases.
\begin{figure*}
\vspace{1.5cm}
\caption{Log$_{10}$ of the NIR light-concentration index $C_{31}$ vs. the
absolute $H$-band magnitude (left panel), and vs.  T,
the index of stage along the Hubble sequence from the RC3 (right panel).
Solid symbols are used for galaxies hosting
active nuclei (Seyferts, LINERS, and starbursts).
The horizontal
lines identify the $C_{31}$ values of pure exponential disks (see text).
}
\label{conc_abstyp}
\end{figure*}

The effective diameter $D_e$ reported in Col. 12 of Table 3 is the major axis 
in arcmin of the fitted elliptical isophote containing 
half the flux corresponding
to the total magnitude $H_T$. 
The average $D_e$ uncertainty, not including the error on $H_T$ is about 2\%
but can be worse, up to 10\%, in case of peculiarly disturbed  morphologies.
\subsection{Concentration indices}
The concentration index $C_{31}$, Col. 13 of Table 3,
is defined as the ratio 
between the major axes of the elliptical isophotes enclosing 75\%
and 25\% of the flux corresponding to $H_{21.5}$ (Gavazzi et al. 1990).
We have explored the possibility of biases in the measurement of $C_{31}$, by 
checking for correlations of the index with apparent magnitude, apparent size, and ellipticity,
and found no evidence of any.
As noted by Gavazzi et al. (1996a), $C_{31}$ depends on the galaxy absolute 
luminosity 
and, to a lesser degree, on morphological type;
such relations, for the present sample, are illustrated in 
Fig. \ref{conc_abstyp}.
As was the case also in Gavazzi et al. (1996a), 
the highest concentration indices
are found in the most luminous Sb spirals. 
It also appears from Fig. \ref{conc_abstyp}
that the relation between 
$\log C_{31}$ and NIR absolute magnitude is not simply 
linear but definitely concave
or L shaped. That is to say that, not only the mean value, 
but also the variance of
the concentration index increases with luminosity. 
In Fig. \ref{conc_abstyp} we indicate the value of $C_{31}$
of a pure exponential disk: the upper value, $\log C_{31}\simeq 0.45$ is found
when the 25\% and 75\% fractions refer to the total disk luminosity;
the lower one, $\log C_{31}\simeq 0.41$, when the fractions refer to the
luminosity within 3.5 exponential scale lengths, that is the average value
corresponding to $H_{21.5}$ (see Sect. \ref{magnitudes}).
Not surprisingly, there is a definite tendency of the 
lowest luminosities and latest
morphologies to conform to a pure disk but some low 
concentration indices are present
also among the luminous, early-type objects.
A more detailed study of the
characteristics of the bulge and disk components of these objects is deferred to
forthcoming papers.

Ten of the sample galaxies, that is $\sim$ 6\% of the total,
are reported to host active nuclei, either Seyferts or LINERS or starbursts;
they are marked with solid symbols in Fig. \ref{conc_abstyp}.
While in the present sample they are quite luminous, there is no particular tendency
to high $C_{31}$ values; also the distribution among the morphological types
is rather uniform but for the avoidance at $T\geq 6$.
We find no difference, for $C_{31}$, between the different classes of activity.
\begin{figure*}
\vspace{1.5cm}
\caption{ $\langle\mu_{21.5}\rangle$, 
the average NIR surface brightness within the isophote 
at 21.5 $H$-mag arcsec$^{-2}$ vs. the apparent $H$ magnitude (left panel)
and vs. the isophotal $D_{21.5}$ diameter (right panel).
As in Fig. \ref{radii}, different symbols refer to different ranges 
of optical axial ratios.}
\label{sb_hd}
\end{figure*}

Figure \ref{sb_hd} is a scatter diagram of the average surface brightness
within the 21.5 $H$-mag arcsec$^{-2}$ isophote versus $H_{21.5}$ and vs.
$D_{21.5}$.
While there is a definite correlation between  $\langle\mu_{21.5}\rangle$
and $H_{21.5}$, 
in the sense that faint surface brightnesses are preferentially 
observed in the faintest sample objects,
it disappears almost completely between  $\langle\mu_{21.5}\rangle$
and $D_{21.5}$. The correlation is therefore determined
by the limited range of the selected diameters (see Sect. \ref{select})
rather than by intrinsic properties.
As for the isophotal diameters, the trend of  $\langle\mu_{21.5}\rangle$
with inclination is not significant.
\subsection{\protect\label{ell_sect}Ellipticities and Position Angles}
The results of the elliptical-isophote fitting have been
used to compute estimates of ellipticity $\epsilon$ 
and position angle $P.A.$ 
of the outer regions. 
Our $\epsilon$ and $P.A.$ values are computed 
directly from the output of the IRAF-STSDAS 
routine ``ellipse'' and are the weighted average of the three outermost
points where the ellipticity is still evaluated with a precision 
better than 0.1. The position angles are not reported for galaxies
nearly face-on and particularly uncertain cases are flagged with a `` : ''.
We estimate the $1\sigma$ uncertainty of the ellipticity to be
$\sim 0.05$, in well-behaved cases. 

It turns out that such a blind, although objective,
procedure is often inaccurate.
Indeed, an inspection of the radial profiles
of $\epsilon$ and $P.A.$ in Fig. 4 shows that they are often
determined by the geometry of the spiral pattern,
which often dominates even in the NIR, rather than
by the effective orientation of the disk. This is true in particular for
late spirals seen nearly face on and, obviously, for the more
disturbed, peculiar morphologies. 
In addition our images have a restricted field of view and
are somewhat shallower than the plates
from which the optical values were estimated and therefore
our estimate of the outer disk can be, in some cases, rather
uncertain.
As a consequence, the values we derive
sometimes deviate considerably from those reported in the catalogues,
which are also shown for comparison in Fig. 4.
A direct comparison of our ellipticities, $\epsilon_H$, and those
from the RC3, $\epsilon_B$, is shown in Fig. \ref{ell}. 
As expected the scatter is rather
large and not appreciably influenced by the most uncertain values;
we count 25 galaxies out of 174 for which 
$|\epsilon_H - \epsilon_B| > 0.2$ .
It is also quite clear that most discrepant values are found for
low-inclination objects, where $\epsilon_H$ tends to be definitely
larger than $\epsilon_B$. 

\begin{figure}
\vspace{1.5cm}
\caption{The NIR ellipticity of the outer isophotes $\epsilon_H$
vs. $\epsilon_B$, the ellipticity listed in optical catalogues.
Solid points refer to particularly uncertain $\epsilon_H$ values.
The dashed line is for the case $\epsilon_H = \epsilon_B$;
the dotted lines enclose the region $|\epsilon_H - \epsilon_B| < 0.2$ .}
\label{ell}
\end{figure}
\begin{figure}
\vspace{1.5cm}
\caption{Elliptically averaged radial profile of surface magnitude 
$\mu$, ellipticity $\epsilon$, and position angle $P.A.$ for the
barred Sb UGC 12039. The influence of the bar on the isophotal
fitting between 5 and 20 arcsec is clearly illustrated in all
three profiles. The solid line in the top panel represents the 
surface magnitude profile obtained by imposing at all radii a fixed
$\epsilon = 0.29$ and a fixed $P.A. = 0^\circ$, the average values of the
outer disk
}
\label{u12039}
\end{figure}
\subsection{\protect\label{bars}Non-axisymmetric structures}
The fact that the elliptical-isophote fitting might occasionally be
misled by non-\-axi\-sym\-me\-tric struc\-ture is bound to bear an influence 
on the parameters derived by the brightness profile. For example,
a strong bar on a very tenuous disk
will induce artificially high values of the ellipticity;
the result of such narrow isophotes will be larger isophotal diameters
and consequently incorrect estimates of $H_{21.5}$ and $C_{31}$.
To check the impact of such an effect we have selected the galaxies for which
our estimates of $\epsilon$ and $P.A.$ differ most from the catalogued values
and have repeated the fitting procedure by imposing on the outer regions
the $\epsilon$ values from the RC3. In most cases, as  expected, 
such optical values favour rounder images with a
consequent shrinking of $D_{21.5}$. The effect, in any case, 
is not dramatic and confined to less than 10\%, although it reaches
 over 30\% in 
the two most deviant cases: UGC 1471 and 1626. The effect on 
$H_{21.5}$ is in the sense of making it brighter at
lower ellipticity, but rarely exceeds the errors associated
with noise and calibration. Also the influence on $C_{31}$ is 
appreciable only in the worst cases and even here within
10\%; lower ellipticities tend to yield lower concentration indices.

In a certain number of cases, the isophotal fitting, 
although strongly influenced by 
non-\-axi\-sym\-me\-tric struc\-tures in the inner regions,
is able to
recover the actual $\epsilon$ and $P.A.$ of the outer disk.
These cases ($\sim 20$) 
are characterized
by sudden and strong jumps in their $\epsilon$ and $P.A.$ profiles.
A good example is UGC 12039, which has both a strong bar and
strong spiral arms. As shown in Fig. \ref{u12039}, the
radial profiles of $\mu$, $\epsilon$, and $P.A.$ all show an
obvious jump at $\sim 20$ arcsec, that is right after the 
fading of the bar. In the same figure we also show the $\mu$
profile obtained by imposing a fixed $\epsilon$ and $P.A.$ value,
0.29 and $0^\circ$ respectively, which are the average
values for the outer regions.
The influence of the bar on the derived surface brightness profile
is clearly depicted, with the strongest deviations reaching $\sim 0.5$ mag.
However, in these cases,
the global photometric parameters
(magnitudes, diameters, indices) estimated in the two
ways are virtually identical,
with differences well within the limits of the quoted
accuracy. 
For uniformity with the rest of the sample,
and given the small bearing on the global parameters, we prefer to
retain also in case of strong asymmetries the fitting procedure
with free $\epsilon$ and $P.A.$. In turn this yields the
observed $\epsilon$ and $P.A.$ profiles,
together with higher order azimuthal Fourier components of the
luminosity distribution (Carter 1978), which will be used in
a forthcoming paper of this series to investigate the
properties of bars and non-\-axi\-sym\-me\-tric struc\-tures  
in general.

%
%
\section{\protect\label{concl}Summary}
We have presented the results of a programme of NIR
imaging of disk galaxies in the Pisces-Perseus supercluster area,
$22^h<R.A.<4^h$ and $0^\circ<\delta<45^\circ$.
This programme constitutes the first large NIR imaging survey of the
supercluster and complements previous similar surveys in the Coma and
Virgo areas.
The present paper is the first of a series and contains images
in the $H$ passband of a composite sample of 174 galaxies.

\begin{enumerate}
\item
The sample galaxies,
a subset of the PP supercluster galaxies, are mainly late spirals with a median 
$T\sim 5$.
The sample starts 
lacking completeness at $B_T = 14.0$,
and extends up to $B_T = 16.5$ . 
\smallskip
\item
For each galaxy we present a grey-scale image reaching an isophote of 
$\sim 21.0\, H$-mag arcsec$^{-2}$. For each image we supply 
the elliptically-averaged radial profile of the surface brightness, the
ellipticity, and the position angle of the isophotes.
\smallskip
\item
We derive a total magnitude $H_T$, an isophotal magnitude
$H_{21.5}$, an effective diameter $D_e$, an isophotal one $D_{21.5}$,
the concentration index $C_{31}$, and estimate the ellipticity 
$\epsilon$ and the position angle $P.A.$ of the outer disk. These data,
together with basic information regarding the sample objects and their
observation, are collected in a single table which is available in electronic 
form at the CDS. 
Radial profiles in tabular form and
images FITS files are also available upon request.
\smallskip
\item
The trends of $(H_{21.5}-H_T)$ with $H_{21.5}$ and 
$\langle\mu_{21.5}\rangle$ imply that the average outer disk has an isophotal
radius $D_{21.5}/2 \simeq 3.5 r_d$ and a central surface brightness
$\mu (0) \simeq 17.5$ $H$-mag arcsec$^{-2}$.
\smallskip
\item
The empirical relation in our sample, $D_{21.5} = 0.89 D_{25}$, implies
that the $B-H$ color of the outer disk is bluer, on average, than 3.5,
in agreement with de Jong (1996). The ratio $D_{21.5}/D_{25}$ depends,
although weakly, on the total galaxy $B-H$, and consequently on
its absolute magnitude. 
\smallskip
\item
We confirm the dependence of the (model independent)
concentration index $C_{31}$ on luminosity and, to a lesser degree, on
morphological type. Large values for $C_{31}$, that is values in excess 
of what is expected for pure disks ($\log C_{31}\ga 0.5$), 
are preferentially observed in galaxies
brighter than $M_H\simeq -23$ and $T\la 3$.
\smallskip
\item
Even in the NIR,
objective estimates of outer ellipticities and position angles can
often be contaminated by the geometry of bars and spiral patterns;
the effect is strongest for shallow images,
and for low inclinations.
Their impact on the measurement of global
photometric properties is, in any case, not dramatic.
It virtually affects only $D_{21.5}$, with discrepancies exceeding 10\%
only in the two worst cases.
\end{enumerate}

{\bf Acknowledgements:} G.M. warmly acknowledges the hospitality of
the Astronomy Dept. of Cornell University, where part of the data reduction
was done. We acknowledge the partial support of the Italian
Space Agency (ASI) through the grant ARS--98--116/22. 

{}
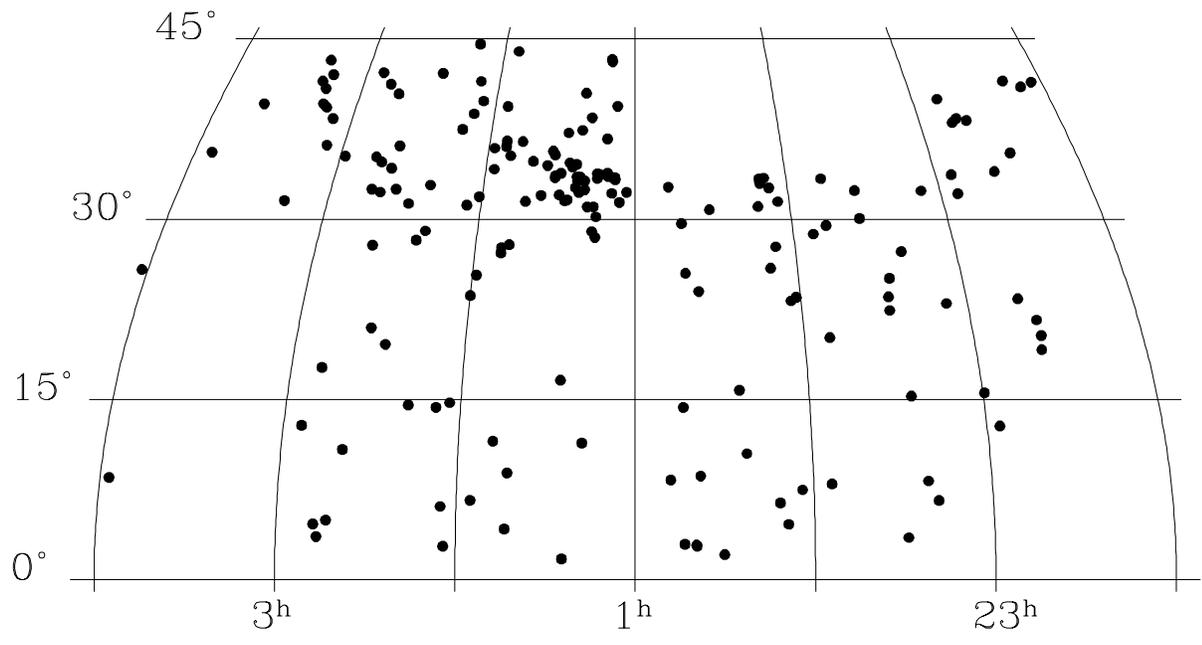
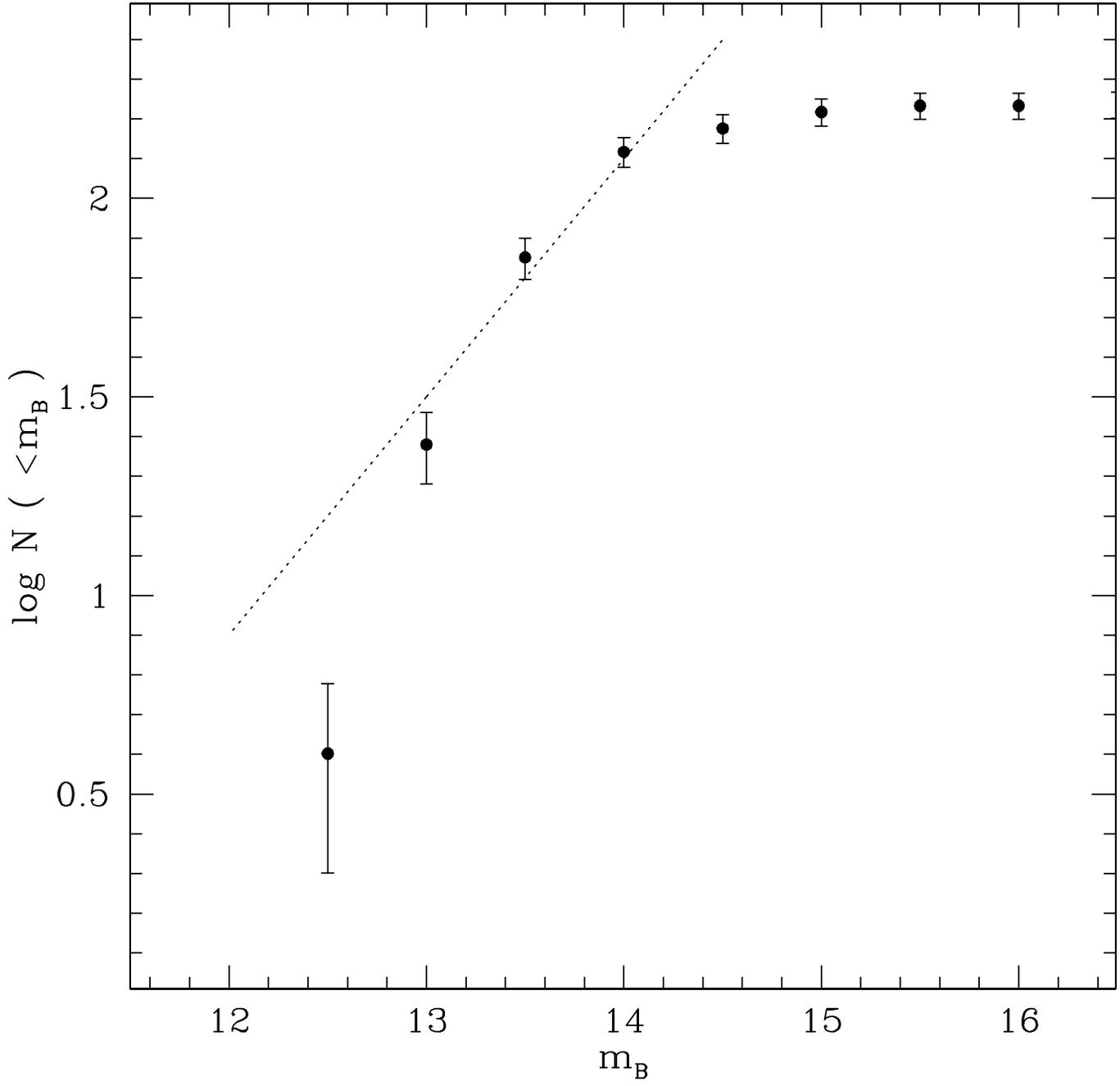
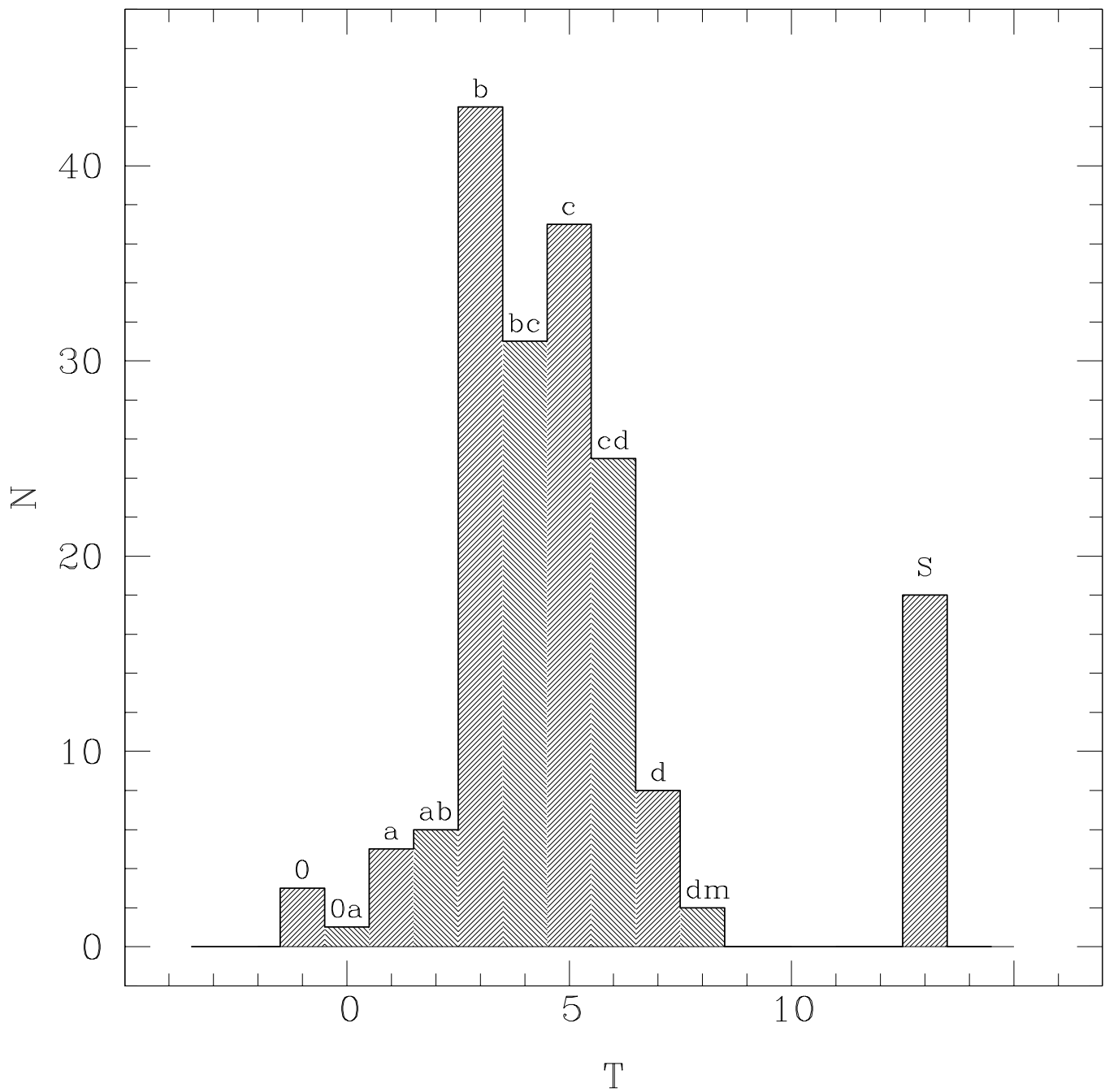
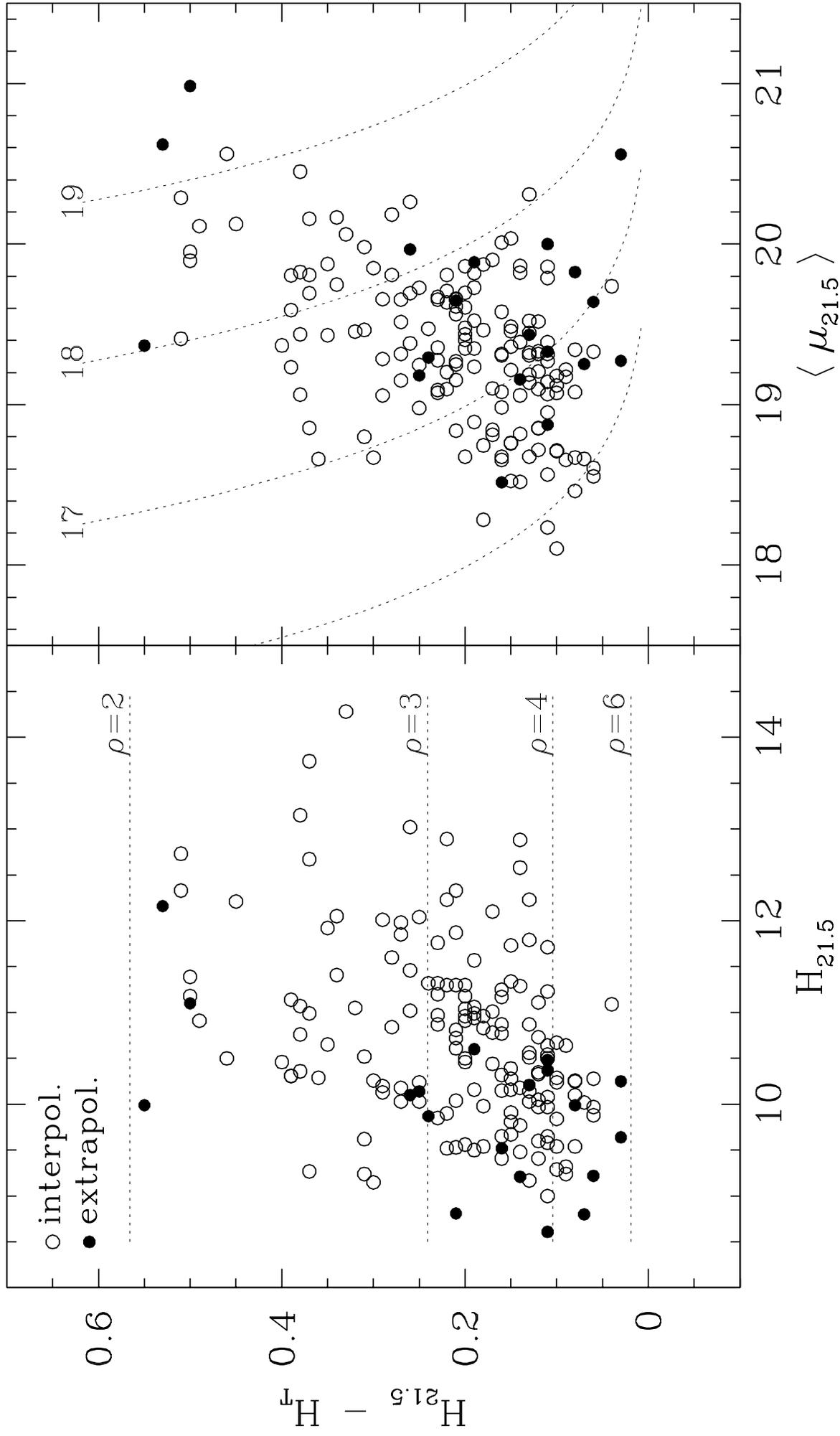
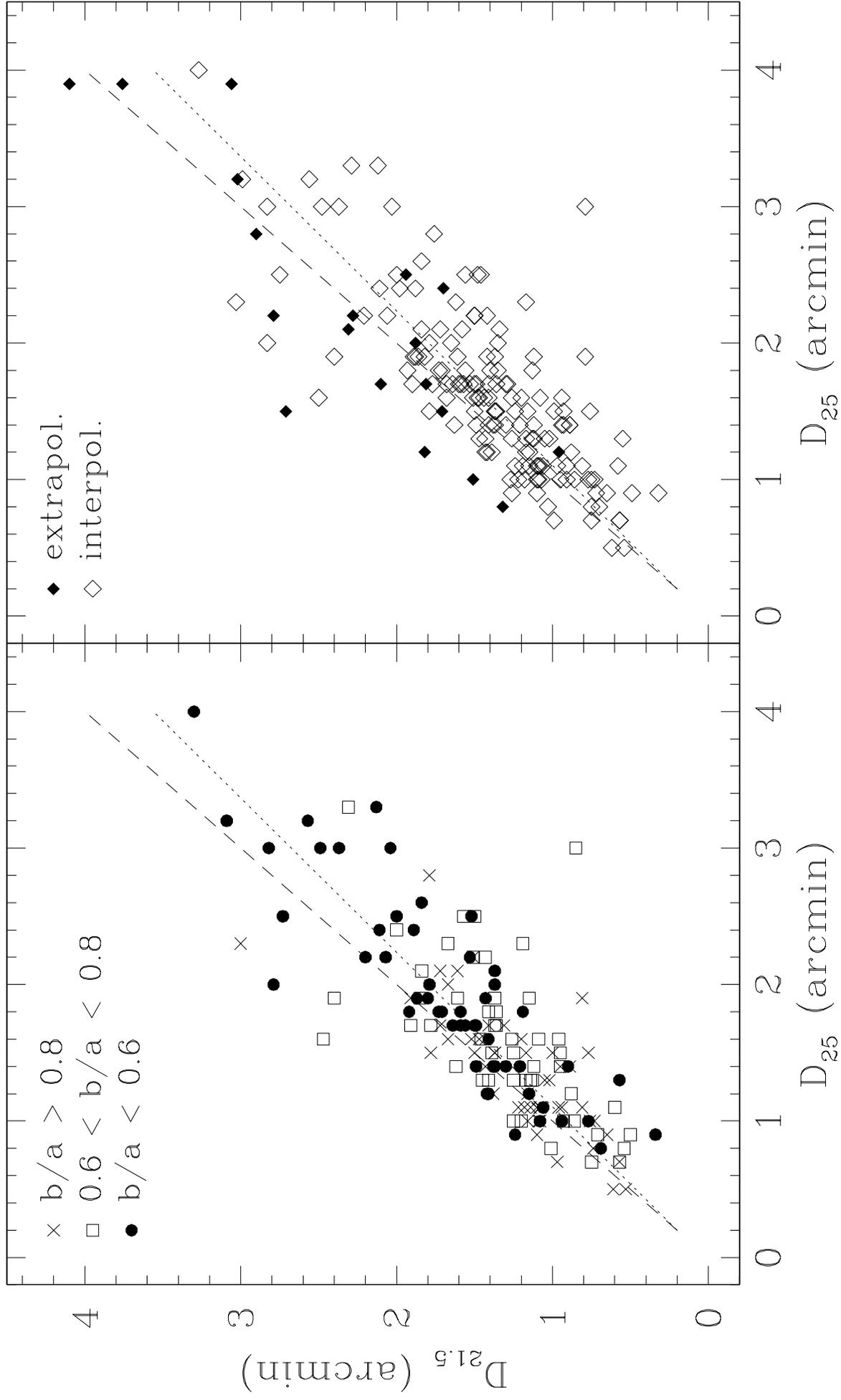
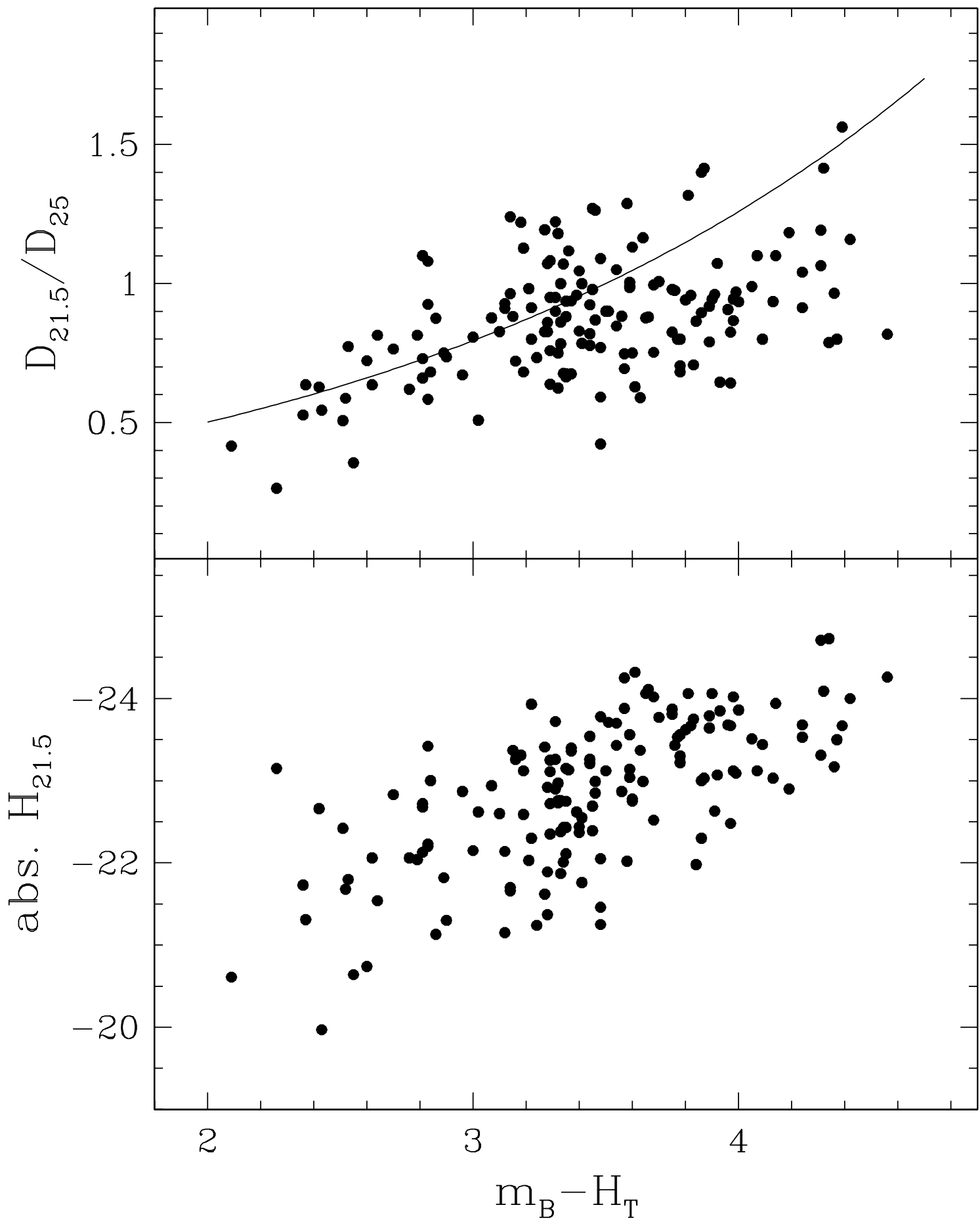
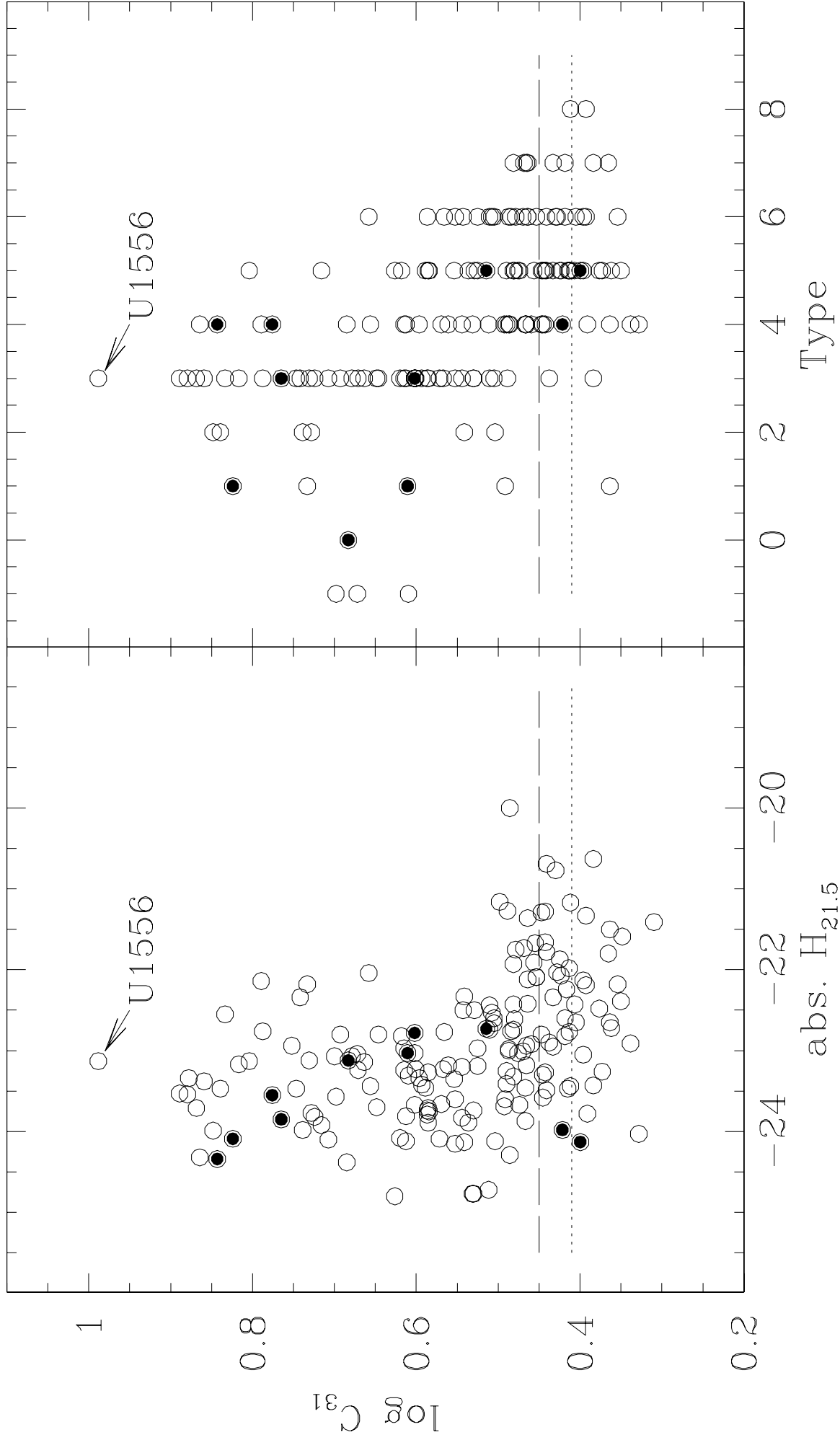
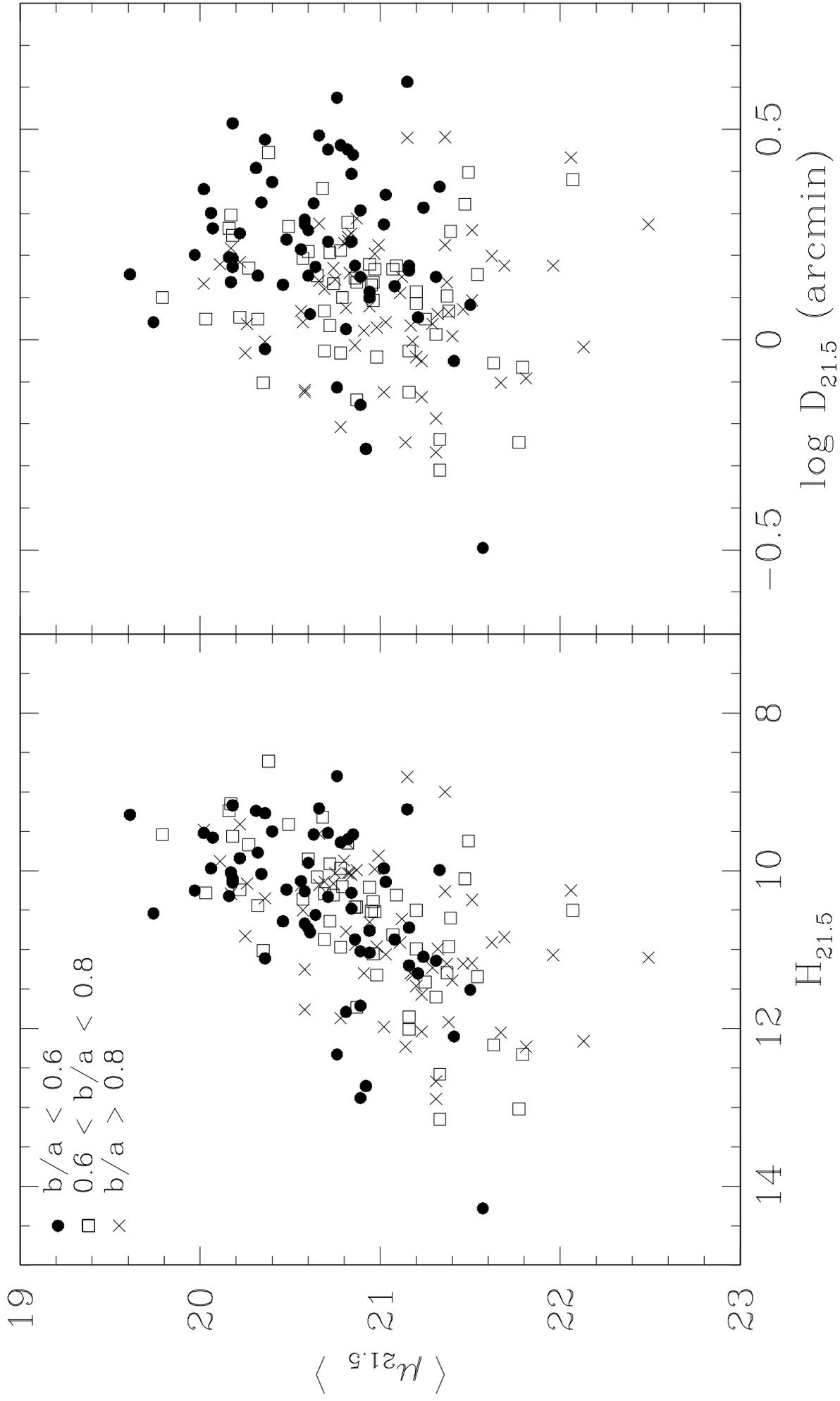
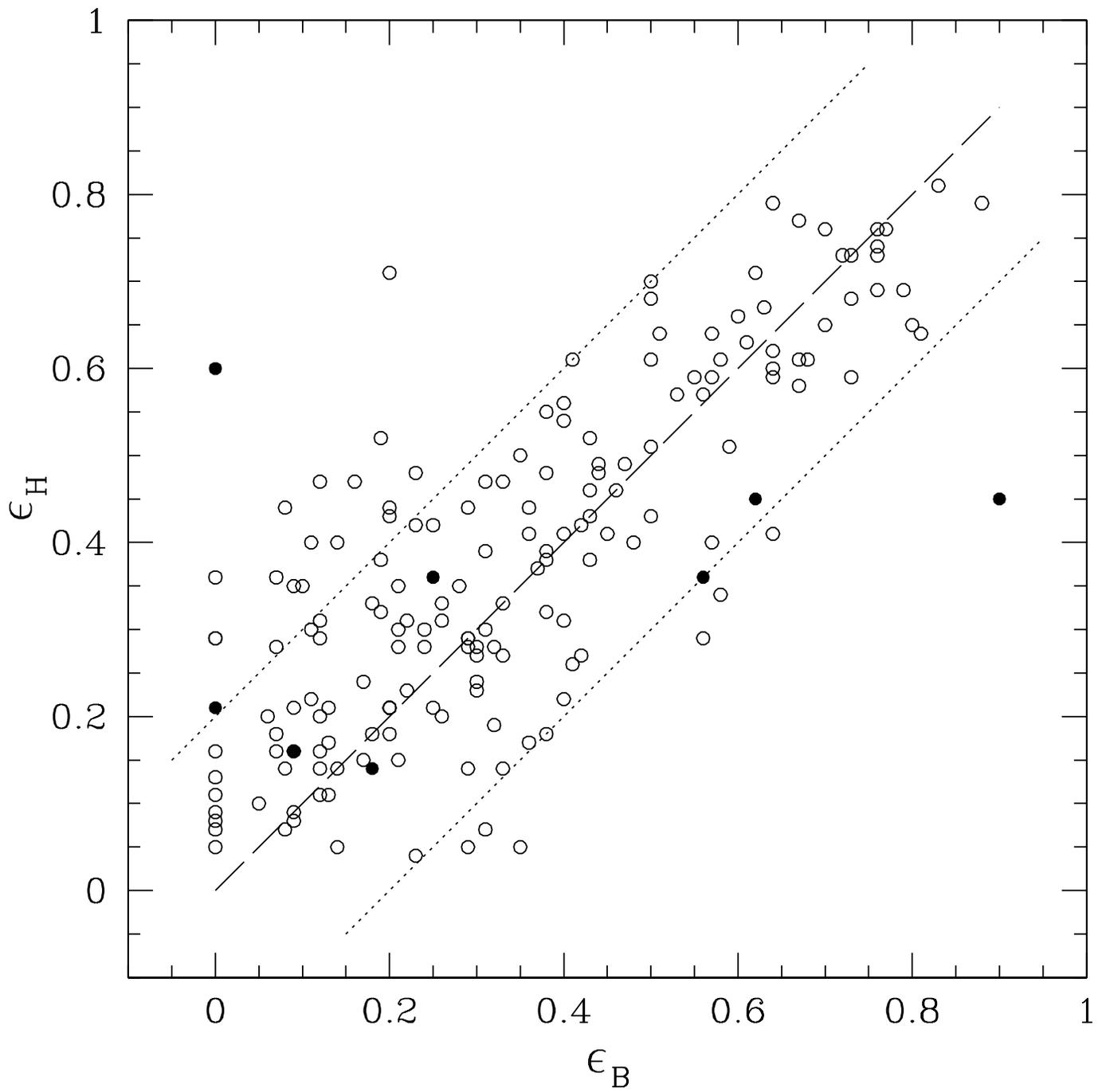
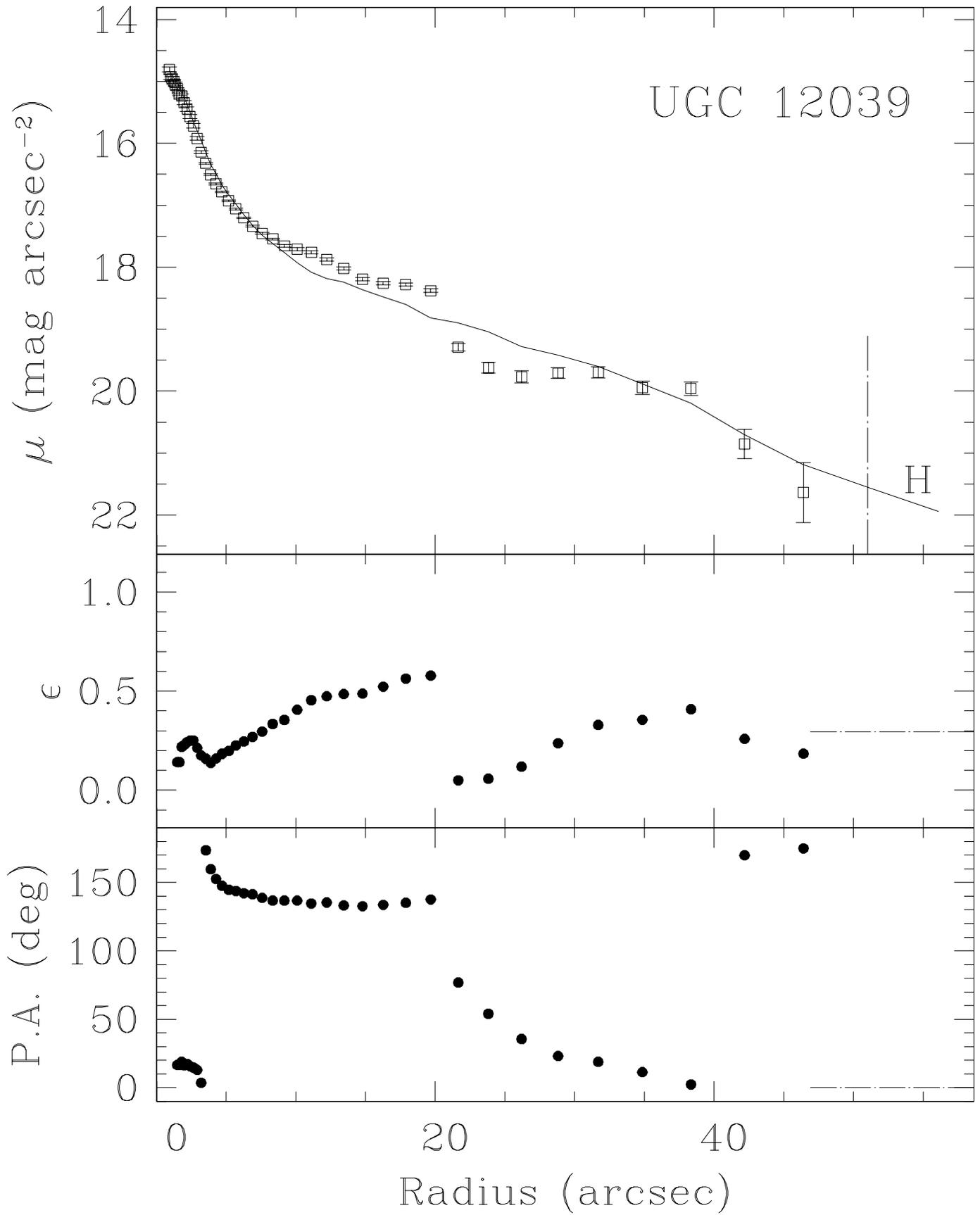
\end{document}